\documentclass[a4paper]{jpconf}
\usepackage{graphicx}
\usepackage{graphics}
\usepackage{iopams}
\usepackage[all]{xy}

\newcommand{\half}{\mbox{$\textstyle \frac{1}{2}$}}

\newcommand{\threehalf}{\mbox{$\textstyle \frac{3}{2}$}}
\newcommand{\re}{\mbox{$\rm e$}}
\newcommand{\ri}{\mbox{$\rm i$}}
\newcommand{\rd}{\mbox{$\rm d$}}

\begin{document}
\title{Entanglement of three-qubit geometry}

\author{Dorje C Brody${}^1$, Anna C T Gustavsson${}^2$ and Lane P
Hughston${}^3$}

\address{${}^1$Department of Mathematics, Imperial College, London
SW7 2AZ, UK}

\address{${}^2$Blackett Laboratory, Imperial College, London
SW7 2AZ, UK}

\address{${}^3$Department of Mathematics, King's College London,
The Strand, London WC2R 2LS, UK}


\begin{abstract}
Geometric quantum mechanics aims to express the physical
properties of quantum systems in terms of geometrical features
preferentially selected in the space of pure states. Geometric
characterisations are given here for systems of one, two, and
three spin-$\half$ particles, drawing attention to the
classification of quantum states into entanglement types.
\end{abstract}

\section{Introduction}

In this article we sketch how the entanglement properties of
elementary spin systems can be described in a geometric language.
The geometric formulation of quantum mechanics has its origin in
the work of Kibble~\cite{Kibble}, and has been developed by many
authors (see references cited in \cite{gqm}). The idea is as
follows. The space of pure states is the space of rays through the
origin of Hilbert space. In particular, the expectation of an
observable ${\hat H}$, given by $\langle\psi|{\hat
H}|\psi\rangle/\langle\psi|\psi\rangle$, is invariant under the
transformation $|\psi \rangle \to \lambda |\psi\rangle$, $\lambda
\in \mathbb{C}\backslash\{0\}$. The ray space thus obtained is the
complex projective space $\mathbb{P}^{n}$, where $n+1$ is the
complex dimension of the associated Hilbert space. Each quantum
state, or Dirac `ket' vector $|\psi\rangle$, projects to a point
in $\mathbb{P}^{n}$. The totality of states $|\eta\rangle$
orthogonal to $|\psi\rangle$ projects to form a hyperplane
$\mathbb{P}^{n-1}$ of codimension one, obtained by solving the
linear equation $\langle{\bar\psi} |\eta\rangle=0$. If
$\mathbb{P}^{n}$ is the space of projective `ket' vectors, then
the aggregate of hyperplanes in $\mathbb{P}^{n}$ is the dual space
of projective `bra' vectors; thus we obtain a \emph{Hermitian
correspondence} between points and hyperplanes.

Let $\{\xi^{\alpha}\}_{\alpha=0,\ldots,n}$ denote the components of
the ket $|\xi\rangle$. Then $\xi^{\alpha}$ can be regarded as the
\emph{homogeneous coordinates} of the corresponding point in
$\mathbb{P}^{n}$. If $\xi^{\alpha}$ and $\eta^{\alpha}$ represent a
pair of distinct states, then the set of all possible superpositions
of these states, given by $\psi^{\alpha} = a \xi^{\alpha} + b
\eta^{\alpha}$, where $a,b \in {\mathbb C}$, projectively
constitutes a \emph{complex projective line} $\mathbb{P}^{1}$.

The quantum state space has a natural Riemannian structure given by
the Fubini-Study metric. The transition probability between a pair
of states $\xi^{\alpha}$ and $\eta^{\alpha}$ is determined by the
associated geodesic distance on $\mathbb{P}^{n}$: $\cos^2
\half\,\theta = \xi^{\alpha} \bar{\eta}_{\alpha} \eta^{\beta}
\bar{\xi}_{\beta}/\xi^{\gamma} \bar{\xi}_{\gamma} \eta^{\delta}
\bar{\eta}_{\delta}$. Conversely, we can \textit{derive} the
Fubini-Study metric from the transition probability~\cite{hughston}.
To see this we set $\theta=\rd s$, $\xi^{\alpha}=\psi^{\alpha}$, and
$\eta^{\alpha}=\psi^{\alpha} + \rd\psi^{\alpha}$, Taylor expand to
second order each side of the expression for the transition
probability, and obtain the line element $\rd s^2 =
4(\bar{\psi}_{\gamma}\psi^{\gamma})^{-2}(
\bar{\psi}_{\alpha}\psi^\alpha \rd \bar{\psi}_{\beta} \rd \psi^\beta
- {\bar\psi}_\alpha \psi^{\beta}\rd{\bar\psi}_{\beta} \rd
\psi^\alpha)$.

Another important feature of the quantum state space is its
\textit{symplectic structure}. The unitary evolution in Hilbert
space is represented by the Hamilton equation on $\mathbb{P}^{n}$,
where the Hamiltonian function is given by the expectation of the
energy operator ${\hat H}$. The metrical geometry of
$\mathbb{P}^{n}$ thus captures probabilistic aspects of quantum
mechanics, and the symplectic geometry of $\mathbb{P}^{n}$
describes the dynamical aspects of quantum mechanics. When these
two are put together, we obtain a fully geometric characterisation
of quantum theory.

\section{A spin-$\half$ particle}
\label{sec:2}

We consider first the case of a single spin-$\half$ particle. The
Hilbert space is $\mathbb{C}^2$, spanned by a pair of spin
eigenstates corresponding to the eigenvalues, say, $S_z=\pm\half$.
The spin-$z$ eigenstates can be written $|\!\uparrow\rangle$ and
$|\!\downarrow\rangle$, and a generic state $|\psi\rangle$ is thus
$|\psi\rangle=a |\!\uparrow\rangle+ b|\!\downarrow\rangle$, $a,b
\in {\mathbb C}$. Projectively, the state space is a line
$\mathbb{P}^{1}$. In real terms this is a two-sphere of radius
$\half$. To see this we recall that a general mixed state of the
spin-$\half$ particle is represented by the density matrix:
\begin{eqnarray}
{\hat\rho} = \left( \begin{array}{cc} t-z & x-\ri y \\ x+\ri y &
t+z \end{array} \right),
\end{eqnarray}
where the trace condition $\tr {\hat\rho}=1$  implies $t=\half$.
Writing $r^2=x^2+y^2+z^2$, we find that the eigenvalues of
${\hat\rho}$ are $\lambda_\pm=t\pm r$. Since ${\hat\rho}$ is
nonnegative, the eigenvalues are nonnegative: $t-r\geq 0$. The trace
condition then says that $x^2+y^2+z^2\leq(\half)^2$. In other words,
the space of $2\times2$ density matrices is a ball of radius
$\half$. For pure states, the density matrix is degenerate with
$\lambda_-=0$; that is, $x^2+y^2+z^2=(\half)^2$. Hence the pure
state space is the surface of the ball.

\begin{figure}[th]
\begin{center}
\includegraphics[width=14pc]{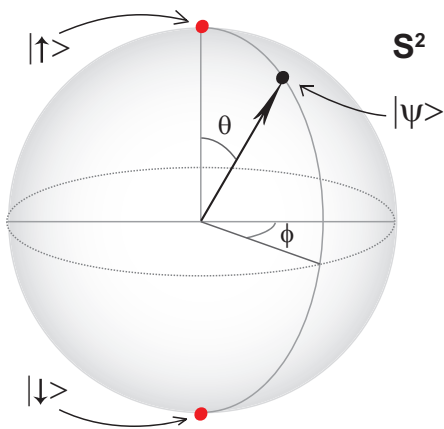}\hspace{2pc}%
\begin{minipage}[b]{20pc}
\caption{\label{S2} Isomorphism between the state space
$\mathbb{P}^{1}$ of a spin-$\half$ system and the two-sphere
$\mathbb{S}^2$. A state $|\psi\rangle$ in $\mathbb{P}^{1}$
corresponds to a point on $\mathbb{S}^2$ and hence can be expressed
in the form $|\psi\rangle\! =\! \cos\half\,\theta
{|\!\uparrow\rangle} + \sin\half\, \theta \, \re^{{\rm
i}\phi}{|\!\downarrow\rangle}$. Since a two-sphere $\mathbb{S}^2$
can be embedded in $\mathbb{R}^3$, we can use the isomorphism
$\mathbb{P}^{1}\sim\mathbb{S}^2$ to identify the north pole of the
sphere as the $S_z=\frac{1}{2}$ `up' state ${|\!\uparrow\rangle}$,
and the south pole as the $S_z=-\frac{1}{2}$ `down' state
${|\!\downarrow\rangle}$. This relation in quantum mechanics is
called the \textit{Pauli correspondence}. Hence we can associate
spin directions with points in the state space.}
\end{minipage}
\end{center}
\end{figure}

A general pure state can be represented by spherical coordinates on
$\mathbb{S}^2$, as shown in Figure~\ref{S2}. We can then use a
two-component spinor notation for quantum states. Letting the spinor
$\psi^A$, $A=0,1$, represent a point on $\mathbb{P}^{1}$, we relate
this to the corresponding point on $\mathbb{S}^2$ by writing
\begin{eqnarray}
\psi^A= \left( \cos \half\,\theta\, , \ \ \sin \half\,\theta\,
\re^{{\rm i}\phi}\right) \qquad (0 \leq \theta \leq \pi,\ 0 \leq
\phi < 2\pi). \label{eq:3}
\end{eqnarray}
We let $\alpha^A=(1,\,0)$ represent the spin-up state (the north
pole on $\mathbb{S}^2$, for which $\theta=0$), and $\bar{\alpha}^A
=\epsilon^{AB} \bar{\alpha}_B=(0,\,1)$ the spin down state, where
$\epsilon^{AB}$ is the anti-symmetric spinor and ${\bar\alpha}_B$ is
the complex conjugate of $\alpha^B$. We use $\alpha^A$ and
$\bar{\alpha}^A$ as our basis in $\mathbb{P}^{1}$ and express the
general pure state as $\psi^A = u \alpha^A + v \bar{\alpha}^A$,
where $(u,v)$ are the homogeneous coordinates of that point.

\section{Two-qubit entanglement}
\label{sec:3}

How is quantum entanglement represented in geometric terms? If one
system is represented by the Hilbert space ${\mathbb C}^{n+1}$, and
another by ${\mathbb C}^{m+1}$, then the combined system is
represented by the tensor product ${\mathbb C}^{n+1} \otimes
{\mathbb C}^{m+1}$. Projectively, the state spaces are given,
respectively, by $\mathbb{P}^{n}$ and $\mathbb{P}^{m}$, while the
state space of the combined system is $\mathbb{P}^{(n+1) (m+1)-1}$.
A characteristic feature of complex projective space is that it
admits what is called the \emph{Segr\'e embedding}: $\mathbb{P}^{n}
\times \mathbb{P}^{m} \hookrightarrow \mathbb{P}^{(n+1)(m+1)-1}$.
The product space $\mathbb{P}^{n} \times \mathbb{P}^{m}$ for the
disentangled quantum states thus `lives' inside the large state
space.

We consider the state space $\mathbb{P}^{3}$ of a pair of entangled
spin-$\half$ particles (the two-qubit system), and represent a
generic state by a spinor $\psi^{AB}$. The disentangled states lie
on the ruled surface $\mathbb{P}^{1} \times \mathbb{P}^{1}
\subset\mathbb{P}^3$, which is the quadric surface $\mathcal{Q}$
given by the solution to the equation
$\epsilon_{AC}\epsilon_{BD}\psi^{AB}\psi^{CD}=0$. States on
$\mathcal{Q}$ are of the form $\psi^{AB}=\psi^A_{1}\psi^B_{2}$ where
each of the spinors $\psi^A_{1}$ and $\psi^B_{2}$ describes one of
the spin-$\half$ particles. The two spinors need not be the same,
and we are free to measure the two spins in different directions.
Hence for the disentangled two-particle states we can write
\begin{eqnarray}
\psi_{1}^A= \left( \cos \half\,\theta_1, \ \ \sin \half\, \theta_1
\, \re^{{\rm i}\phi_1} \right) \quad {\rm and} \quad \psi_{2}^A=
\left( \cos \half\,\theta_2, \ \ \sin \half\, \theta_2\, \re^{{\rm
i}\phi_2}\right). \label{phi^A_1-phi^A_2}
\end{eqnarray}
Here $(\theta_k,\phi_k)$ fix the spin directions of particles
$k=1,2$ on the corresponding Bloch balls.

\begin{figure}[t]
\includegraphics[width=17pc,height=11.5pc]{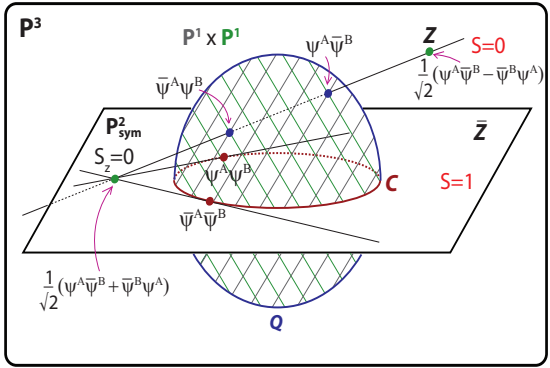}\hspace{2pc}%
\begin{minipage}[b]{19pc}
\caption{\label{fig:2} The state space of two spin-$\half$
particles. The disentangled states form a ruled surface ${\mathcal
Q}=\mathbb{P}^{1}\times\mathbb{P}^{1}$. The singlet state $Z$ in
$\mathbb{P}^{3}$ is invariant under local unitary transformations.
Its conjugate $\bar{Z}$ is the plane $\mathbb{P}^{2}$ of totally
symmetric states, and is spanned by the triplet states given by
$S_z=0,\pm1$. The intersection of ${\mathcal Q}$ and ${\bar Z}$ is a
conic ${\mathcal C}$. The line joining the singlet state and the
spin-$0$ triplet state intersects ${\mathcal Q}$ at two points
corresponding to $|\!\uparrow\downarrow\rangle$ and
$|\!\downarrow\uparrow \rangle$.}
\end{minipage}
\end{figure}

If the system is in a state of total spin zero, it is given by the
totally antisymmetric singlet $Z$ expressible in the form
$\psi^{AB}= \frac{1}{\surd2} \epsilon^{AB}$. The conjugate of the
singlet state is the plane $\mathbb{P}^{2}_{\rm sym}$ of totally
symmetric states. The triplet states with $S^2=1$ and $S_z=1,0,-1$
lie on $\mathbb{P}^{2}_{\rm sym}$ (see Figure~\ref{fig:2}). The
plane of symmetric states intersects the quadric in a conic curve
$\mathcal{C} = \mathcal{Q} \cap \mathbb{P}^{2}_{\rm sym}$. The conic
is generated by a Veronese embedding of $\mathbb{P}^{1}$ in
$\mathbb{P}^{2}_{\rm sym}$, in such a way that the Pauli
correspondence for the spin directions in ${\mathbb R}^3$ is induced
in the state space of higher spins. Each point on $\mathcal{C}$
represents a spin-one state $\psi^{AB}$ for some direction
$(\theta,\phi)$ in ${\mathbb R}^3$. Thus $\mathcal{C}$ has the
topology ${\mathbb S}^2$. The choice of the spin direction fixes the
$S_z=1$ triplet state $\psi^{AB}=\psi^A\psi^B$. Its complex
conjugate is a line which is tangent to the conic at $\psi^{AB}=
\bar{\psi}^A \bar{\psi}^B$, corresponding to the $S_z=-1$ state. The
intersection of the conjugate lines of the two states $S_z=\pm 1$
gives the $S_z=0$ state $\psi^{AB}= \frac{1}{\surd2}
(\psi^A\bar{\psi}^B + \bar{\psi}^B \psi^A)$.

If the system is initially in the singlet state $Z$, or, more
generally, in a superposition of the singlet state and the $S_z=0$
triplet state, there are two disentangled states that can result as
a consequence of a spin measurement along the $z$-axis for one of
the particles. These can be formed by connecting the singlet state
and the $S_z=0$ triplet state with a line. This line intersects the
quadric in two points $\psi^{AB}= \psi^A\bar{\psi}^B$ and
$\psi^{AB}= \bar{\psi}^A\psi^B$, which are the possible results of a
spin measurement.

\section{Three-qubit entanglement}
\label{sec:4}

\begin{figure}
\begin{center}
\includegraphics[width=0.75\textwidth]{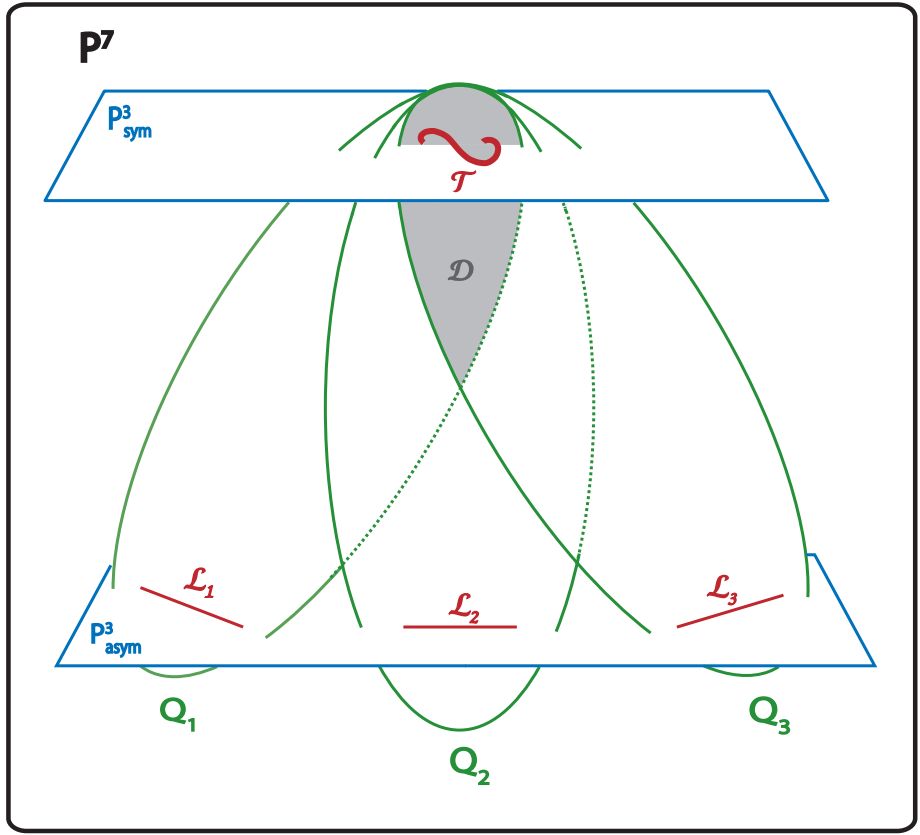}
\caption{\label{fig:3} The state space of three spin-$\half$
particles. There are three configurations of partly entangled states
$\{Q_i\}_{i=1,2,3}$. Their intersection is the space of totally
disentangled states $\mathcal{D}$. The symmetric states form a
hyperplane $\mathbb{P}^{3}_{\rm sym}$ whose orthogonal complement
$\mathbb{P}^{3}_{\rm asym}$ intersects each $Q_i$ in a line
$\mathcal{L}_i$. The line ${\mathcal L}_i$ represents for each $i$
the state space of particle $i$ when the two remaining particles are
entangled to form an $S=0$ singlet.}
\end{center}
\end{figure}

The geometry of the three-qubit system is very rich. The state space
$\mathbb{P}^{7}$ of the three-qubit system is obtained by projecting
the tensor product space ${\mathbb C}^2\otimes {\mathbb C}^2\otimes
{\mathbb C}^2$. There are several different types of entanglement
that can result~\cite{sergio}. First, we have the completely
disentangled states. These constitute a triply-ruled three-surface
$\mathcal{D}= \mathbb{P}^{1} \times \mathbb{P}^{1} \times
\mathbb{P}^{1}$. Next, we have the partly entangled states, where
one of the particles is disentangled from the other two. There are
three such systems of partly entangled states, each of which
constitutes a four-dimensional variety $Q_i \subset{\mathbb P}^7$
($i=1,2,3$) with the structure of $\mathbb{P}^{1} \times
\mathbb{P}^{3}$. The ${\mathbb P}^1$ in each case represents the
state space of the disentangled particle, and the ${\mathbb P}^3$
represents the state space of the remaining entangled pair. It
should be evident that $\mathcal{D}=Q_1\cap Q_2\cap Q_3$. Finally,
we have the states for which all three particles are entangled.
Figure~\ref{fig:3} shows a schematic illustration of
$\mathbb{P}^{7}$, highlighting the totally disentangled and
partially entangled state spaces.

The states of total spin $S=\frac{3}{2}$ form a hyperplane ${\mathbb
P}^3_{\rm sym}\subset{\mathbb P}^7$. These states are represented by
totally symmetric spinors, i.e. those satisfying
$\psi^{ABC}=\psi^{(ABC)}$, where the round brackets denote
symmetrisation. The states of total spin $S=\frac{1}{2}$ also
constitute a hyperplane of dimension three, which we call ${\mathbb
P}^3_{\rm asym}$. The `asymmetric' states are those that are of the
form
\begin{eqnarray}
\psi^{ABC}=\alpha^A\epsilon^{BC} +\beta^B\epsilon^{CA}+
\gamma^C\epsilon^{AB}
\end{eqnarray}
for some $\alpha^A, \beta^A,\gamma^A$. It should be evident that the
symmetric states and the asymmetric states are orthogonal. Thus
${\mathbb P}^3_{\rm asym}$ is the orthogonal complement of ${\mathbb
P}^3_{\rm sym}$ in ${\mathbb P}^7$.

The hyperplane ${\mathbb P}^3_{\rm sym}$ intersects ${\mathcal D}$
in a twisted cubic curve $\mathcal{T} = \mathbb{P}^{3}_{{\rm sym}}
\cap \mathcal{D}$. See \cite{wood} for the properties of the twisted
cubic. This curve is given by the Veronese embedding of
$\mathbb{P}^{1}$ in $\mathbb{P}^{3}_{\rm sym}$, which takes the form
$\psi^A\to\psi^A\psi^B\psi^C$ (see, e.g., references
\cite{gqm,EHH}). The Veronese embedding induces the Pauli
correspondence in the state space $\mathbb{P}^{7}$. In particular,
if $S=\frac{3}{2}$, then we have a quadruplet of possible spin
states relative to the $z$-axis, with $S_z= \threehalf, \half,
-\half,-\threehalf$. The hyperplane $\mathbb{P}^{3}_{{\rm sym}}$ is
spanned by $S=\threehalf$ quadruplet. In particular, the
$S_z=\threehalf$ and $S_z=-\threehalf$ states lie on $\mathcal{T}$.
If the spinor $\psi^A$ has $S_z=\frac{1}{2}$, then the
$S_z=\threehalf$ and $S_z=-\threehalf$ states are given by
$\psi^{ABC}=\psi^A \psi^B \psi^C$ and $\psi^{ABC}=\bar{\psi}^A
\bar{\psi}^B \bar{\psi}^C$, respectively.

\begin{figure}[h]
\includegraphics[width=18pc,height=13.0pc]{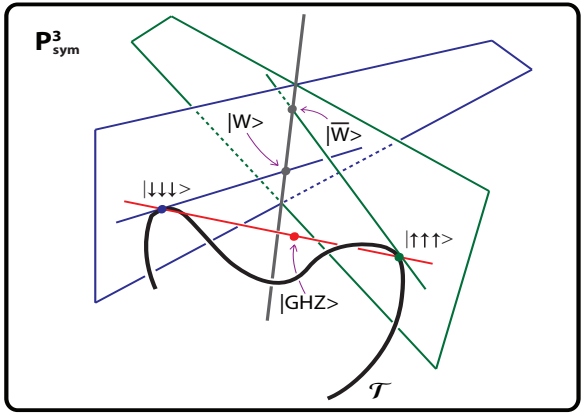}\hspace{2pc}%
\begin{minipage}[b]{18pc}
\caption{\label{fig:4} A close up of the hyperplane
$\mathbb{P}^{3}_{{\rm sym}}$ of symmetric states. The twisted cubic
$\mathcal{T}$ is the intersection between this hyperplane and the
space of totally disentangled states $\mathcal{D}$. For every spin
direction the states ${|\!\uparrow\uparrow\uparrow\rangle}$ and
${|\!\downarrow\downarrow\downarrow\rangle}$ lie on the twisted
cubic. The GHZ state $|\textrm{GHZ}\rangle =\frac{1}{\surd2}
({|\!\uparrow\uparrow \uparrow\rangle} +
{|\!\downarrow\downarrow\downarrow\rangle})$ is found on the line
joining them, and the $S_z=\frac{1}{2}$ Werner state
$|\textrm{W}\rangle = \frac{1}{\surd{3}}({|\!\uparrow\uparrow
\downarrow\rangle} + {|\!\uparrow\downarrow \uparrow\rangle} +
{|\!\downarrow \uparrow\uparrow\rangle})$ is given by the
intersection of the osculating plane at ${|\!\uparrow\uparrow
\uparrow\rangle}$ and the tangent line at
${|\!\downarrow\downarrow\downarrow\rangle}$.}
\end{minipage}
\end{figure}

The tangent line to a point $\alpha^A\alpha^B\alpha^C$ on ${\mathcal
T}$ consists of spinors of the form $\psi^{ABC}=\alpha^{(A}\alpha^B
x^{C)}$ for some $x^A$. The so-called osculating 2-plane at
$\alpha^A \alpha^B \alpha^C$ consists of spinors of the form
$\psi^{ABC}= \alpha^{(A}x^B y^{C)}$ for some $x^A,y^A$. Clearly, the
tangent line lies on the osculating plane. The two-dimensional
envelope generated by the tangent lines to ${\mathcal T}$ generates
a quartic surface ${\mathcal H}_{\rm sym}$ in ${\mathbb P}^3_{\rm
sym}$, given by the equation $Q_{AB}Q^{AB}=0$, where
$Q_{AB}=\psi_A^{\ CD}\psi_{BCD}$ and $\psi^{ABC}=\psi^{(ABC)}$. The
states of ${\mathbb P}^3_{\rm sym}$ are of three types: those on
${\mathcal T}$; those on ${\mathcal H}_{\rm sym}\backslash{\mathcal
T}$; and those on ${\mathbb P}^3_{\rm sym}\backslash{\mathcal
H}_{\rm sym}$. The complex conjugate of a point $\psi^A\psi^B\psi^C$
on $\mathcal{T}$ is a six-dimensional hyperplane in $\mathbb{P}^7$,
which intersects $\mathbb{P}^{3}_{{\rm sym}}$ at the osculating
plane of the point ${\bar\psi}^A{\bar\psi}^B{\bar\psi}^C$ on
$\mathcal{T}$. As illustrated in Figure~\ref{fig:4}, the
intersection of the tangent line of $\mathcal{T}$ at the
$S_z=\threehalf$ state and the osculating plane of $\mathcal{T}$ at
the $S_z=-\threehalf$ state is the $S_z=\frac{1}{2}$ state
$\frac{1}{\surd3}(\psi^A \psi^B \bar{\psi}^C + \psi^A \bar{\psi}^B
\psi^C + \bar{\psi}^A \psi^B \psi^C)$. Similarly, the
$S_z=-\frac{1}{2}$ state $\frac{1}{\surd3} (\psi^A \bar{\psi}^B
\bar{\psi}^C + \bar{\psi}^A \psi^B \bar{\psi}^C + \bar{\psi}^A
\bar{\psi}^B \psi^C)$ is the intersection of tangent line of
$\mathcal{T}$ at the $S_z=-\threehalf$ state and  the osculating
plane of $\mathcal{T}$ at the $S_z=\threehalf$ state.

The space $\mathbb{P}^{3}_{\rm asym}$ intersects each of the
varieties $\{Q_i\}_{i=1,2,3}$ in a line $\mathcal{L}_i = Q_i \cap
\mathbb{P}^{3}_{\rm asym}$. These lines represent the singlet states
of the entangled pairs in the $\{Q_i\}$. This follows because the
states of $\mathbb{P}^{3}_{{\rm asym}}$ can be expressed in the form
$\alpha^A\epsilon^{BC}+\beta^B\epsilon^{CA}+ \gamma^C\epsilon^{AB}$.
The intersection of $\mathbb{P}^{3}_{{\rm asym}}$ with $Q_1$ thus
takes the form $\alpha^A\epsilon^{BC}$, where
$\alpha^A\in\mathbb{P}^{1}$ and $\epsilon^{BC}\in\mathbb{P}^{3}$.
Since $\epsilon^{BC}$ is antisymmetric, it corresponds to the
singlet state in $\mathbb{P}^{3}$. Then as $\alpha^A$ varies, we
obtain the `singlet' line $\mathcal{L}_1$. This configuration is
illustrated in Figure~\ref{fig:3}.

An interesting feature of 3-qubit entanglement is that under
stochastic local operations and classical communication (SLOCC)
operations there are six different equivalence classes of
entanglement~\cite{DurVidalCirac}. The SLOCC operations are elements
of the group $SL(2,\mathbb{C})^{\otimes 3}$. The space of
equivalence classes is then $\mathbb{C}^2 \otimes \mathbb{C}^2
\otimes \mathbb{C}^2 / SL(2,\mathbb{C})^{\otimes 3}$. Two states are
equivalent under SLOCC if there exists an invertible local operation
interpolating them. The six classes are the totally disentangled
states, the three configurations of partly entangled states, and the
two different classes of totally entangled states: those that are
locally equivalent to the Greenberger-Horne-Zeilinger state $|{\rm
GHZ}\rangle= \frac{1}{\surd2}(\psi^A \psi^B \psi^C + \bar{\psi}^A
\bar{\psi}^B \bar{\psi}^C)$, and those locally equivalent to the
Werner state $|{\rm W} \rangle= \frac{1}{\surd{3}}(\psi^A \psi^B
\bar{\psi}^C + \psi^A \bar{\psi}^B \psi^C + \bar{\psi}^A \psi^B
\psi^C)$.

The GHZ state is symmetric and lies on the chord that joins the two
quadruplet states that lie on the twisted cubic. The GHZ state is
usually said to be the maximally entangled state, in the sense that
it maximally violates the Bell inequalities. If one of the particles
is disentangled from the rest, however, then the other two are
automatically disentangled. The W state on the other hand maximises
2-qubit entanglement inside the 3-qubit state, so that if one of the
particles is disentangled it leaves the other two maximally
entangled. The `three-tangle' $4|{\rm Det}(\psi)|$, where ${\rm
Det}(\psi)$ is the Cayley hyperdeterminant~\cite{Gelfand}, is zero
for all states except those states that are equivalent to the GHZ
state (see \cite{Miyake,levay}). The idea is as follows. We consider
the space ${\mathcal D}$ of totally disentangled states, and let
${\mathcal H}$ denote the six-dimensional variety generated by the
system of 3-hyperplanes tangent to ${\mathcal D}$. Then ${\mathcal
H}$ turns out to be a quartic surface in ${\mathbb P}^7$, consisting
of those points for which the Cayley invariant vanishes. In
particular, ${\mathcal H}$ is given by
\begin{eqnarray}
\psi_A^{\ BC}\psi_{BCD} \psi_P^{\ RS} \psi_{QRS}
\epsilon^{AP}\epsilon^{BQ}=0.
\end{eqnarray}
A necessary and sufficient condition for $\psi^{ABC}$ to satisfy
this relation is that
\begin{eqnarray}
\psi^{ABC} = x^A \beta^B \gamma^C + \alpha^A y^B \gamma^C + \alpha^A
\beta^B z^C \label{eq:4}
\end{eqnarray}
for some $\alpha^A, \beta^B, \gamma^C, x^A, y^B, z^C$. In
particular, if $\alpha^A\beta^B\gamma^C$ is a point in ${\mathcal
D}$, then the tangent plane to ${\mathcal D}$ at that point consists
of states of the form (\ref{eq:4}) for some choice of $x^A, y^B,
z^C$. On the other hand, under the SLOCC classification, the states
that are equivalent to the GHZ state have nonzero hyperdeterminant.
This can be seen from the fact that the GHZ state lies on a chord of
${\mathcal T}$. In particular, we note that ${\mathcal H}_{\rm sym}
={\mathcal H}\cap{\mathbb P}^3_{\rm sym}$.

There are other open questions in describing entanglement, both
geometrically and algebraically. For example, is there a
geometrically unambiguous measure of pure-state entanglement for the
3-qubit system? If so, can it be extended to $n$-qubits for $n\geq
4$? Right now there is an active search being undertaken to find a
good way of quantifying the amount of entanglement in a quantum
state. We hope a geometric formulation will provide intuitive
answers to these questions.

\ack DCB acknowledges support from The Royal Society. ACTG would
like to thank the organisers of the DICE2006 conference in Piombino,
Italy, 11-15 September 2006 for the opportunity to present this
work. 
\end{document}